\documentclass[]{iopart}

\usepackage{iopams}
\usepackage{graphicx}
\usepackage{braket}

\newcommand{\dvec}[1]{\ensuremath{\boldsymbol{#1}}}

\newcommand{\vA}{\dvec{\mathrm{A}}}
\newcommand{\vE}{\dvec{\mathrm{E}}}

\newcommand{\vk}{\dvec{\mathrm{k}}}
\newcommand{\vp}{\dvec{\mathrm{p}}}

\newcommand{\vr}{\dvec{\mathrm{r}}}

\newcommand{\kVcm}{\ensuremath{\mathrm{kV}\,\mathrm{cm}^{-1}}}
\newcommand{\THz}{\ensuremath{\,\mathrm{THz}}}
\newcommand{\fa}{\mathfrak{a}}

\begin{document}

\title{Irradiated bilayer graphene}
\author{D. S. L. Abergel$^{1,2}$ and Tapash Chakraborty$^2$}
\address{$^1$ Condensed Matter Theory Center, University of Maryland, College
Park, MD USA 20742.}
\address{$^2$ Department of Physics and Astronomy, University of
Manitoba, Winnipeg, MB, Canada, R3T 2N2.}

\begin{abstract}
	We describe the gated bilayer graphene system when it is subjected
	to intense terahertz frequency electromagnetic radiation. We examine
	the electron band structure and density of states via exact
	diagonalization methods within Floquet theory.
	We find that dynamical states are induced which lead to modification
	of the band structure. We first examine the situation where there is
	no external magnetic field. 
	In the unbiased case, dynamical gaps appear in the spectrum which
	manifest as dips in the density of states. For finite interlayer
	bias (where a static gap is present in the band structure of
	unirradiated bilayer graphene), dynamical states may be induced in
	the static gap. These states can show a high degree of valley
	polarization.
	When the system is placed in a strong magnetic field, the radiation
	induces coupling between the Landau levels which allows dynamical
	levels to exist. For strong fields, this means the Landau levels are
	smeared to form a near-continuum of states.
\end{abstract}

\maketitle

\section{Introduction \label{sec:intro}}

Graphene \cite{novoselov-sci306} and its bilayer \cite{ohta-sci313}
have attracted much attention recently due to the novel fundamental
physics that they display and huge scope that they have for device
applications \cite{chakraborty-advphys,bonaccorso-natphoton4}. 
In particular, electrons in monolayer graphene (whose low-energy
quasiparticles mimic chiral Dirac fermions with linear dispersion) have
displayed relativistic-like phenomena, including Klein tunneling
\cite{stander-prl102,young-natphys5} and the half-integer quantum Hall
effect \cite{novoselov-nat438,zhang-nat438}.
Electrons in bilayer graphene show properties which are hybrid
between the monolayer and traditional two-dimensional semiconductor
systems, since the low-energy quasiparticles are chiral, but the
inter-layer coupling induces an effective mass and corresponding
quadratic energy dispersion.
A fully-tunable gap can be opened at the charge neutrality point by
application of a transverse electric field (either by
gating \cite{oostinga-natmat7} or by doping \cite{ohta-sci313}), a feature
which is unique to this system.

Monolayer graphene which is irradiated by monochromatic, high-intensity
laser light has been studied theoretically for the bulk
\cite{vasko-prb77, satou-prb78, rodriguez-prb78, mikhailov-jpcm20,
wright-apl95, oka-prb79}, in nanoribbons \cite{liu-apl93, qiang-prb79},
and in $np$-junctions \cite{fistul-prl98, syzranov-prb78}. Experimental
investigations of few-layer graphene have also been carried out
\cite{choi-apl94}. 
However, the transport properties of bilayer graphene which is
irradiated by intense laser light have not been considered in much
detail. Ryzhii \textit{et al}. proposed that a phototransistor could be
implemented using bilayer graphene \cite{ryzhii-prb79}, Wright
\textit{et al}. have found a large peak conductance in the terahertz and
far infra-red frequency range for bilayer graphene nanoribbons
\cite{wright-prl103}, and we have previously suggested that
valley-polarized electrons can be produced in gapped bilayer graphene
\cite{abergel-apl95}.

In this paper, we give a comprehensive description of the band structure
and density of states of irradiated bilayer graphene, both in zero
external magnetic field and in a strong field. 
This is of fundamental physical importance, but also has application in
the realm of devices and technology because of the growing consensus
that graphene and its bilayer have vast potential in the fields of
optoelectronics and photonics \cite{bonaccorso-natphoton4}, and in the
design of new electronic devices such as ambipolar transistors
\cite{yang-acsnano4}. 
Also, the spin-like degrees of freedom in graphene (such as the lattice
pseudospin and the valley) may allow for electronic implementations of
the ideas of spintronics \cite{zutic-rmp76} which have been discussed in
the literature \cite{abergel-apl95, rycerz-natphys3, xiao-prl99,
martin-prl100, garcia-pomar-prl100}. In all of these areas, a thorough
understanding of the basic properties of irradiated bilayer graphene is
an essential building block for design and application of devices.
In particular, this present work focuses on the generation of
valley-polarized states which may be used as a filter for the generation
of valley-polarized currents. This is an essential step in the
realization of valley-tronics devices.

We briefly outline the structure of our paper. In Section
\ref{sec:model} we describe the theoretical framework which we employ
for both the zero-field and strong-field cases. Then we present and
discuss the results of our calculations in Section \ref{sec:results}
before summarizing and placing our work in the context of valley-tronic
devices in Section \ref{sec:summary}. Various important formulae and
derivations are collected in the Appendix.

\section{Theoretical framework \label{sec:model}}

The Hamiltonian of irradiated bilayer graphene is written as
\begin{equation*}
	\mathcal{H}(t) = H_0 + H_U + h(t)
\end{equation*}
where $H_0$ is the continuum limit of the tight binding Hamiltonian,
$H_U$ describes the external electrostatic field due to a top gate or
dopants, and $h(t)$ is the time-dependent part which depends only on the
irradiating field. 
We take the nearest-neighbour approximation of this $H_0$, but in
principle any combination of hopping terms can be included by computing
the eigenfunctions of the static Hamiltonian numerically.
The static part of the Hamiltonian $H_0 + H_U$ determines single
particle wave functions $\varphi_X$, which span the spatial part of the
Hilbert space of solutions of $\mathcal{H}(t)$ and which we shall use as
a basis for the time-dependent solutions $\Psi(t)$ of $\mathcal{H}(t)$.
The energy spectrum and wave functions of the static Hamiltonian are
presented in \ref{app:SP}, both in zero external magnetic field
and when a strong magnetic field quantizes the electron motion into
Landau levels. The solutions of the time-dependent part are found
via the Floquet theorem, which we describe below.

The time-dependent term in the Hamiltonian is not known \textit{a
priori}, so we describe the process by which it is derived.
The irradiating field is represented as a time-dependent vector
potential which is introduced to the static Hamiltonian via Peierl's
substitution $\vp\to\vp + e\vA$, where $e>0$ is the magnitude of the
electron charge.  We write the vector potential as $\vA(t) = [A_x(t),
A_y(t)]$, and substituting into the static Hamiltonian in
Equation \eref{eq:H0def} to find
\begin{equation}
	h(t) = \xi v e \sigma_x \otimes
	\left( \begin{array}{cc} 0 & \! A_x(t) - iA_y(t) \\
	A_x(t) + iA_y(t) \! & 0 \end{array} \right)
	\label{eq:Htdef}
\end{equation}
where $\sigma_x$ is the first Pauli matrix. The vector potential encodes
all information about the radiation, such as the polarization of the
light, strength of the field, and frequency of the radiation.
In the case of linearly polarized light, this field can be represented
as $\vA_{\mathrm{lin}} =  A \cos\Omega_A t\left[ \cos\theta, \sin\theta
\right]$ where $\theta$ is the angle of polarization in the plane of the
graphene and $\Omega_A$ is the frequency.
The strength of the vector field is given by the parameter
$A=F/\Omega_A$ where $F = |\vE|$ is strength of the electric field.
However, when the light is circularly polarized, there is a phase
difference between the $x$ and $y$ components of the field so that
\begin{equation}
	\vA_\pm = A \left[ \cos\Omega_A t, \pm\sin\Omega_A t \right]
	\label{eq:circpol}
\end{equation}
where the plus or minus sign refer to right-handed and left-handed
polarizations, respectively. Full details of the time-dependent
Hamiltonians in each case are given in \ref{app:matelems}.

It is not possible to develop exact solutions to the eigenvalue equation
for $\mathcal{H}(t)$. However, since the pertubation $h(t)$ is
periodic, we may employ the Floquet theorem \cite{dittrich-book} to
write the full time-dependent wave function $\Psi(\vr,t)$ as the product
of a periodic function $\Phi(\vr,t)$ (with period $t_0 = 2\pi/\Omega_A$)
and a time evolution function. 
The `temporal Brillouin zone' is the interval $0<t<t_0$, and the wave
function is
\begin{equation}
	\Psi(\vr,t) = e^{-i\varepsilon t/\hbar} \Phi(\vr,t)
	\label{eq:Psidef}
\end{equation}
where $\Phi(\vr,t+t_0) = \Phi(\vr,t)$.
The scalar constant $\varepsilon$ is called the quasienergy. This
theorem is the
temporal analogue of the Bloch theorem, so that the quasienergy is
equivalent to the quasimomentum, and the time period is equivalent
to the lattice constant of the reciprocal lattice. The periodic wave
functions $\Phi$ are called `Floquet states'.
Substituting Equation \eref{eq:Psidef} into the time-dependent
Schr\"odinger equation yields 
$\mathcal{H}(t) \Phi(t) - i\hbar \frac{\partial}{\partial
t}\Phi(t) = \varepsilon \Phi(t)$,
an eigenvalue equation for the operator
$\mathcal{F}(t) = \mathcal{H}(t) - i\hbar \frac{\partial}{\partial
t}$ with $\mathcal{F}(t) \Psi(t) = 0$.

In order to solve the Schr\"odinger equation for $\mathcal{F}(t)$ and
$\Phi(t)$, we consider an expanded Hilbert space
$\mathcal{R}\otimes\mathcal{T}$ of square-integrable functions of
space and functions of time with period $t_0$ (see Ref.
\cite{sambe-pra7} for a full description).
In this space, the scalar product is defined as the regular spatial
scalar product with the average over one period:
\begin{equation*}
	\langle \braket{n|m} \rangle =
	\frac{1}{t_0}\int_0^{t_0}
	\braket{n|m} dt.
\end{equation*}
where $n$ and $m$ label arbitrary states in
$\mathcal{R}\otimes\mathcal{T}$. Since the wave functions are periodic
and the eigenfunctions of the static Hamiltonian $H_0+H_U$ form a
complete set for the spatial
coordinates, we can write the wave functions as
\begin{equation}
	\Phi_m(\vr, t) = \sum_{j,X} c^m_{jX} \bar{\Phi}_{jX}(\vr,t)
	\label{eq:Phidef}
\end{equation}
where $m$ labels the Floquet state and the Hilbert space
$\mathcal{R}\otimes\mathcal{T}$ is spanned by the infinite set of
functions 
\begin{equation*}
	\bar{\Phi}_{jX}(\vr,t) = e^{ij\Omega_A t} \varphi_X(\vr) 
\end{equation*}
such that $j\in \{\ldots,-2,-1,0,1,2,\ldots\}$, and $X$ labels the
eigenstates of $H_0 + H_U$.  The label $X$ contains all appropriate
single-particle quantum numbers, but its exact composition depends on
whether there is a magnetic field present in the system.
The operator in Equation \eref{eq:Htdef} does not couple states with
different momenta, different spin, or which are in different valleys.
Therefore the Floquet states which result from diagonalization of the
$t$-dependent Hamiltonian retain these three quantities as good quantum
numbers.
The Hamiltonian can be written as a matrix by computing the matrix
elements of $\mathcal{F}(t)$ over these states. This yields an
infinite dimensional matrix which can be truncated for a sufficiently
large number of terms in the Fourier expansion and numerically
diagonalized to give quasienergies and wave functions to arbitrary
precision. These matrix elements are discussed in \ref{app:matelems}.

We now introduce the two-time formalism which we use to compute the
Green's function and hence the density of states (DoS) in irradiated
bilayer graphene. 
In this formalism, the time associated with the expanded Hilbert space
(which was previously labelled `$t$' but which we shall call $\zeta$
from now on) is separated from the evolution of the system such that the
full time-dependent solution of the Schr\"odinger equation $\Psi(\vr,t)$
is \cite{dittrich-book}
\begin{equation}
	\Psi(\vr,t) = \Psi(\vr,\zeta,t)\Big|_{\zeta=t}. \label{eq:psittp}
\end{equation}
The two-time wave function is then defined to be
\begin{equation*}
	\Psi(\vr,\zeta,t) = \exp\left(-\frac{i}{\hbar}
	\mathcal{F}(\zeta)(t-t^\prime) \right) \Psi(\vr,\zeta,t^\prime)
\end{equation*}
where $\mathcal{F}$ is the Floquet operator introduced earlier.
Full time-dependent solutions are given by the limiting procedure in
Equation \eref{eq:psittp}, but we shall generally be interested in the
dynamics of the system on timescales much longer than $t_0$ so we shall
instead take the time average with respect to the field.

In the Matsubara formalism, we utilize the grand canonical ensemble, and
define the associated energy scale $\kappa = \varepsilon - \mu$ where 
$\mu$ is the chemical potential. The operator for this energy
is $\mathcal{K} = \mathcal{F} - \mu N$. 
The imaginary time $\tau=it$ is defined and the evolution of the field
operators associated with the Floquet states is given by (the coordinate
dependence is implicit)
\begin{eqnarray}
	\psi(\zeta,\tau) = \sum_n e^{-\kappa_n \tau/\hbar}
		\Phi_n(\zeta) a_n, \\
	\psi^\dagger(\zeta,\tau) = \sum_n e^{\kappa_n \tau/\hbar} 
		\Phi^\ast_n(\zeta) a^\dagger_n,
\end{eqnarray}
where the index $n$ runs over all Floquet states.
The Matsubara Green's function is
\begin{equation*}
	\mathcal{G}(\vr \zeta, \vr' \zeta', \tau-\tau')  = 
	\frac{-1}{\hbar} \mathrm{Tr} \left[
		e^{-\beta(\mathcal{K}-\Omega)} T_\tau \psi(\zeta,\tau)
		\psi^\dagger(\zeta',\tau') \right].
\end{equation*}
The operator $T_\tau$ is the $\tau$-ordering operator, $\Omega$ is the
thermodynamic potential and serves as the normalizing factor for the
thermodynamic average, and $\beta = 1/(k_B T)$.
The Fourier transform of this Green's function is the quantity from which
the density of states can be calculated. The Fourier transform is
\begin{eqnarray}
	\mathcal{G}(\vp,i\omega_n) = \frac{1}{t_0^2}
		\int_0^{t_0} d\zeta \int_0^{t_0} d\zeta'
		\int_0^\beta d(\tau-\tau') e^{i\omega_n \tau} \times \nonumber \\
	\qquad\qquad \qquad\qquad 
	\times \frac{1}{\mathcal{A}}\int d^2\vr \int d^2\vr' 
		e^{i\vp\cdot\vr} e^{-i\vp'\cdot\vr}
	\mathcal{G}(\vr\zeta\tau,\vr'\zeta'\tau')
	\label{eq:GFT}
\end{eqnarray}
which includes an averaging procedure over the period of the fast
oscillation associated with the radiation.
The retarded Green's function $G^R$ can then be found by carrying out
the analytic continuation $i\omega_n \to \omega + i\delta$ and the
density of states can be extracted from this Green's function in the
standard way:
\begin{equation}
	\rho(\omega) = -\frac{1}{\pi} \mathrm{Im} \sum_{\vp}
	\mathrm{tr} \, G^R(\vp,\omega) \label{eq:DoSdef}
\end{equation}
where $\mathrm{tr}$ denotes the summation of the diagonal elements of
${G}^R$ which is a $4\times 4$ matrix in the sublattice space.


\section{Results \label{sec:results}}

\subsection{Zero magnetic field \label{sub:zmf}}

\begin{figure}[tb]
	\centering
	\includegraphics[]{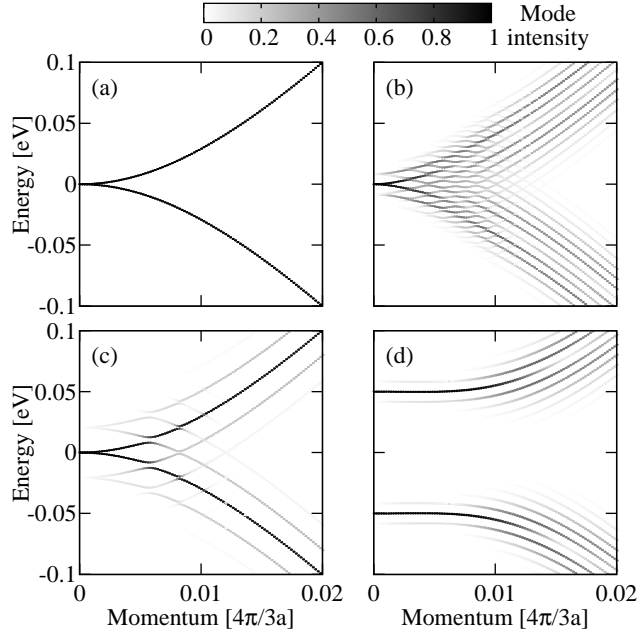}
	\caption{Energy spectrum of irradiated bilayer graphene in zero
	external magnetic field under right-handed circularly polarized
	light. (a) The unirradiated spectrum; (b)
	$\Omega_A = 2\THz$, $F=5\,\kVcm$;
	(c) $U =100\,\mathrm{meV}$, $\Omega_A = 2\THz$, $F=5
	\,\kVcm$;
	(d) $\Omega_A = 5\THz$, $F=5 \,\kVcm$.
	\label{fig:bzerospectrum}}
\end{figure}

In the case of zero external magnetic field, the single particle quantum
states in the system are of plane wave nature and are characterized by a
two-dimensional wave vector $\vk$. In addition, there are two (real)
spins $\sigma$, two inequivalent valleys $\xi$, and four bands within 
each valley (labeled by the conduction or valence band $\nu$ and the
high-energy or split branch $b$). The wave functions associated with
these states are given in \ref{app:zerofield}. The energy
spectrum is found by substituting the matrix elements of the radiation
operator into the Hamiltonian and numerically diagonalizing the
resulting matrix. The spectrum is shown in Figure \ref{fig:bzerospectrum}.
Panel (a) shows the unirradiated low energy bands for comparison
with the other three plots. 
In zero magnetic field, the effective coupling parameter is $v_F e
F/\hbar \Omega_A^2$ which implies that the strongest coupling occurs for
smaller frequencies. This is illustrated by panels (b) and (c) which
show the effects of an irradiating field with $\Omega_A = 2
\THz$ and $5 \THz$ respectively with a field
strength of $F = 5\,\kVcm$. The mixing between
different Fourier harmonics is stronger for the lower frequency, but the
dynamic gap opened by the mixing is larger for the higher frequency.
The dynamic gaps only occur when states from opposite bands mix. States
from the same band run parallel to each other (seperated in energy by
$\hbar\Omega_A$) and therefore cannot cross. There is no gap at $k=0$
because the distribution of the wavefunction across the four lattice
sites forbids coupling for small momenta (see \ref{app:SP}).

\begin{figure}[tb]
	\centering
	\includegraphics[]{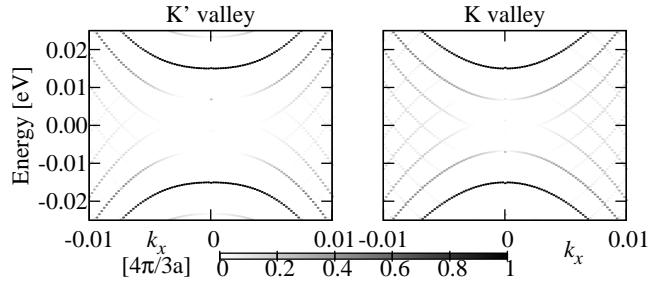}
	\caption{Valley asymmetry when interlayer bias potential is present.
	In these plots $F= 5 \,\kVcm$, $\Omega_A = 2
	\THz$, and $U = 30\,\mathrm{meV}$.
	\label{fig:vpspec}}
\end{figure}

When a gap is introduced to the spectrum by doping one side of the
graphene bilayer or by electrostatic gating, the effect of the radiation
is markedly different. This is because the gap means that consecutive
Fourier harmonics from opposite bands do not cross, and therefore the
spectral weight is spread only between Floquet states which originate
from the same band. Therefore no anticrossings appear, although the
effect of a broadening of the band will be seen. This is illustrated in
panel (d) of Figure \ref{fig:bzerospectrum}. However, as
Figure \ref{fig:vpspec} shows, the coupling to right-handed circularly
polarized light is stronger in the K valley than in the K$^\prime$
valley. This is due to the redistribution of the wave function among the
four lattice sites as a result of the interlayer bias potential which
generates the gap. If the direction of the potential or the orientation
of the irradiating field are changed, then the coupling becomes stronger
in the K$^\prime$ valley instead.
If linearly polarized light is used, the response of electrons in
the two valleys are identical since linearly polarized radiation
can be represented as the sum of two circularly polarized
components.

\begin{figure}[tb]
	\centering
	\includegraphics{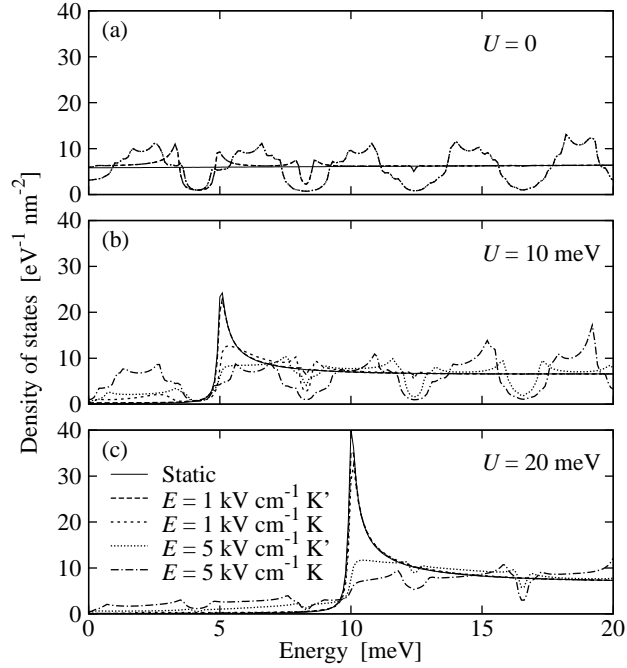}
	\caption{Density of states in bilayer graphene in zero magnetic
	field. (a) For ungapped bilayer graphene, the valley degeneracy
	is intact and gaps open at intervals of $\Omega_A/2 \approx 4.1
	\,\mathrm{meV}.$ (b) $U = 10 \,\mathrm{meV}$. A significant electron
	density is induced in the spectral gap by the radiation. (c) $U =
	20\,\mathrm{meV}$. The spectral gap is too wide to allow
	significant coupling of electron and hole states.
	\label{fig:zfdos}}
\end{figure}

The Floquet wave functions which are calculated numerically can now
be used in Equations \eref{eq:Phidef} and \eref{eq:GFT} to compute
the Green's function and hence the density of states. The Green's
function is
\begin{equation*}
	G^R(\vp,\omega) = \frac{1}{\hbar} \sum_{n'} 
	\frac{1}{\omega - \kappa_{\vp n'}/\hbar +i \delta} \sum_{x} 
	\left| c_{0x}^{\vp n'} \right|^2 
\end{equation*}
where $n'$ labels the discrete set of Floquet states with wave
vector $\vp$ and $x$ labels the unirradiated basis states with wave
vector $\vp$. The density of states can be calculated by
substituting this expression into Equation \eref{eq:DoSdef} and numerically
evaluating the integral over momentum: 
\begin{equation*}
	\rho = \int_0^\Lambda p\,dp
	\sum_{x_\alpha} \frac{\delta}{\left(\hbar\omega-\kappa_\alpha\right)^2 +
	\delta^2} \sum_X \left| c^{p\alpha}_{0X} \right|^2
\end{equation*}
where $\Lambda$ is the momentum cut-off determined by requiring the
summation over all states yields the correct electron density at
half-filling. 
Plots of this function are shown in Figure \ref{fig:zfdos}. In (a), the
interlayer bias potential is zero so that the graphene does not have a
static gap. The density of states for unirradiated graphene is constant
in this case, as predicted by straightforward analysis of the band
structure. When the graphene is irradiated, dynamic gaps open at
$\Omega_A/2$ and are clearly visible for stronger fields. The responses
of electrons in the two valleys are
identical. In (b), the same system parameters are used, except that now
a small static gap is present ($U=10\mathrm{meV} = U/2$). 
In the unirradiated case (thin solid line), the static gap manifests as
a region with near-zero DoS for small energies ($\hbar \omega <
5\mathrm{meV}$). For
$\hbar\omega > 5\mathrm{meV}$, the dynamic gaps are still present.
However, for $\hbar\omega < 5\mathrm{meV}$, a finite DoS is present
under strong radiation. This is due to the dynamical states that are
induced by the radiation. In (c), the gap is wider ($U=20\mathrm{meV}$)
and now the conduction and valence bands are seperated to an extent
where significant coupling between them is not permitted. However,
states within the same band do still couple causing the band edge to be
significantly smeared and for electron density to be present in the
static gap.

\subsection{Quantizing magnetic field \label{sub:fmf}}

\begin{figure}[tb]
	\centering
	\includegraphics{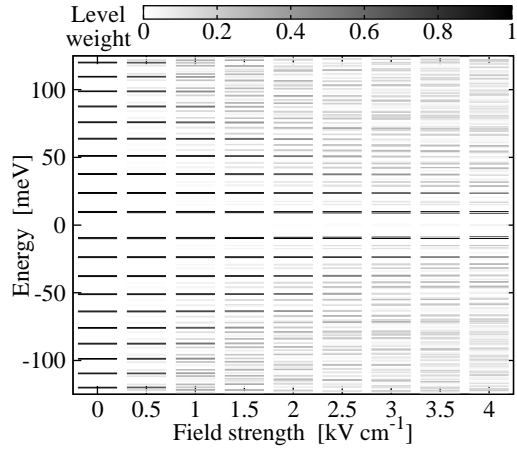}
	\caption{Evolution of energy levels with the strength of a linearly
	polarized irradiating field. The line thickness indicates the
	strength of the mode, $\Omega=1\THz$,
	$U=20\,\mathrm{meV}$, and $B=5\,\mathrm{T}$. The spectrum of
	levels is symmetric about zero energy in this case.
	\label{fig:BEevo}}
\end{figure}

When a strong magnetic field is present in the system, the motion of the
electrons is quantized into Landau levels. Using the Landau guage $\vA_B
= [0,Bx,0]$, the discrete single particle quantum numbers are the same
as in the zero field case, but the momentum is continuous in the $y$
direction and discrete in the $x$ direction. This is due to the gauge
field which breaks the translational symmetry. Therefore, the sum over
the two-dimentional momenta may be split into two separate
one-dimensional sums, one discrete (over $k_x$) and the other
represented by an integral with periodic boundary conditions (over
$k_y$).

Figure \ref{fig:BEevo} shows the evolution of the Landau level spectrum
with increasing intensity of linearly polarized incident radiation. At
$F=0$, the standard bilayer Landau level spectrum is evident. For weak
field, ($F < 1\,\kVcm$) the mixing of dynamical states
is small and the original Landau level spectrum is recognizable. For
strong coupling ($F > 3\,\kVcm$), the Landau level
spectrum is replaced by a near-continuum of levels, each with rather
small weight, and there are the beginings of states visible in the
gapped region. Notice, however, that the two $\delta$-states barely
change their intensity showing that they are only very weakly coupled to
the radiation due to their unique distribution of wave function weight
between the four sublattice sites.

\begin{figure}[tb]
	\centering
	\includegraphics{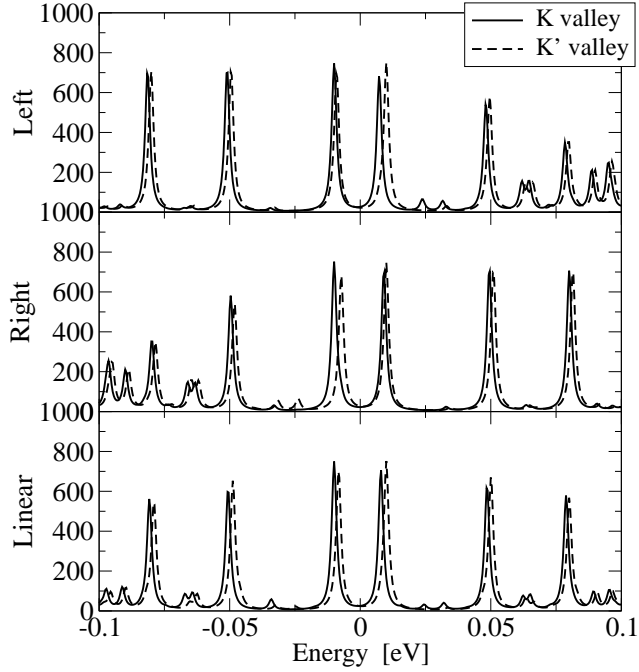}
	\caption{Density of states for $B=12\mathrm{T}$, $\Omega_A =
	4\THz$, $U = 20\mathrm{meV}$, and with $F=5\,\kVcm$ in each
	valley.  The radiation shifts the energy of the Landau level, and
	coupling properties are different in the two bands.
	\label{BDoSU020}}
\end{figure}

The DoS in this case is evaluated using the same steps as in the zero
magnetic field situation. The analytical expression which we must
evaluate for the DoS is
\begin{equation*}
	\rho = \frac{\Lambda}{\pi \lambda_B} \sum_{x_\alpha} 
	\frac{\delta}{\left(\hbar\omega - \kappa_\alpha\right)^2 + \delta^2}
	\sum_X \left| c_{0X}^\alpha \right|^2
\end{equation*}
This function is plotted in Figure \ref{BDoSU020} for bilayer graphene
with a small inter-layer bias $U=20\mathrm{meV}$ for left-handed,
right-handed and linear polarization of the incident light. We clearly
see the radiation-induced dynamical bands as expected.
The first feature of these plots which requires discussion is the
difference between the behavior in the conduction and valence bands for
the two circular polarizations. This occurs because coupling between the
radiation and the light and the electrons depends critically on the wave
function components. The factors like $\xi \Xi \pm \varepsilon$ mean
that the weight on each lattice component are different in the two
bands, implying that the field will affect these electrons differently.
When strong coupling occurs, the Landau level spectrum is almost washed
out by the dynamical bands, but when coupling is weak, the states are
still clearly discernable. There is also a slight shift in the energy of
the Landau levels in each valley, which is caused by the broken valley
degeneracy in the single-particle spectrum [see Equation
\eref{eq:BfieldE}]. The linear polarization shows this as a uniform
downward shift of the Landau levels in the K valley relative to the K'
valley.

\section{Conclusions and summary \label{sec:summary}}

\begin{figure}
	\centering
	\includegraphics[]{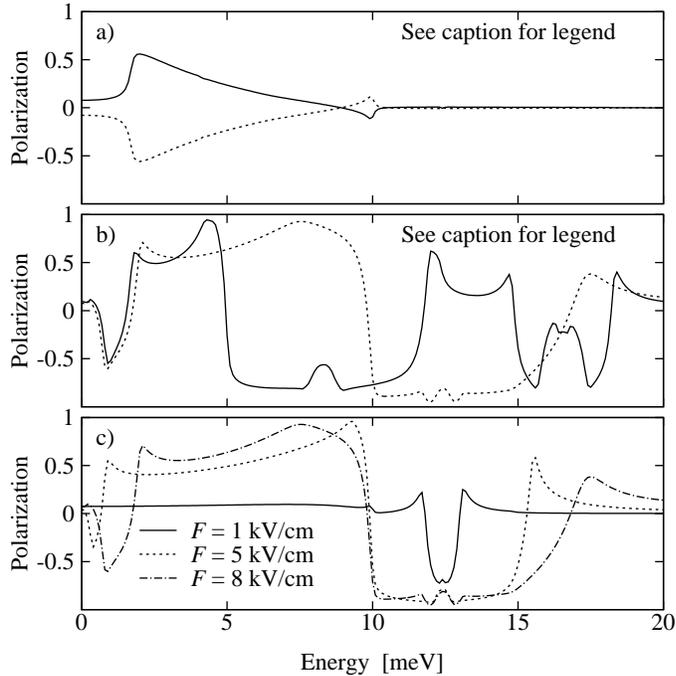}
	\caption{Valley polarization for zero magnetic field. 
	a) Polarization for right-handed (solid line) and left-handed
	(dashed line) radiation with 
	$\Omega_A=2\THz$, $F=1\,\kVcm$, and $U=20\mathrm{meV}$. 
	b) Demonstration that polarization of almost unity can be achieved
	under specific conditions. Solid line has $\Omega_A=4\THz$,
	$F=5\,\kVcm$ and $U=10\mathrm{meV}$; Dashed line
	is $\Omega_A=6\THz$, $F=8\,\kVcm$, and
	$U=20\mathrm{meV}$.
	c) Polarization features caused by the dynamic spectral gaps at
	$\omega = \Omega_A/2$ become more pronouned with increasing field
	stength. In this panel, $\Omega_A = 6\THz$ and
	$U=20\mathrm{meV}$.
	\label{fig:ZerofieldPol}}
\end{figure}

We have shown that under certain conditions, the coupling of electrons
in biased bilayer graphene to the external electromagnetic field is not
the same in either valley.
In order to demonstrate how this might be applicable in the design of a
device, we discuss how the existence of parameter ranges where there is
significant DoS in one valley but not in the other allows us to consider
the possiblity of generating valley-polarized electrons
\cite{abergel-apl95}. 
If a current is incident on a region of irradiated bilayer graphene
where the radiation and inter-layer bias are tuned such that states only
exist in one valley, then incident current in the valley where there are
no states will not be able to traverse the irradiated region. 
This region can then act as a filter for the electron valley, leading to
the possibility of switching devices and `valleytronic' applications
\cite{abergel-apl95}. We define the polarization to be
\begin{equation*}
	\mathcal{P} = \frac{\rho_K - \rho_{K'}}{\rho_K + \rho_{K'}}
\end{equation*}
so that $\mathcal{P}=+1$ implies that all electrons are in the K valley
and $\mathcal{P} = -1$ means that all are in the K' valley.
Figure \ref{fig:ZerofieldPol} shows the polarization for several
different scenarios. Firstly, in panel a) we demonstrate that the
left-handed and right-handed orientations of light induce polarization
in opposite valleys. In panel b) we demonstrate that polarization can
approach unity for specific values of the radiation parameters and bias
potential. This is important if highly-polarized electron currents are
to be produced. Lastly, in panel c) we show that the polarization
induced near the dynamical gaps becomes more pronounced and exists for a
wider energy range when the field strength is increased. This is in
contrast to polarization features caused by additional dynamical states
appearing in the static gap, which exist only for small parameter
ranges. The features associated with the dynamical gaps also become
stronger as the frequency $\Omega_A$ is increased.

In the case where a strong magnetic field is present, valley
polarization may also be generated, but only as a result of the shift in
energy of the Landau levels created by the radiation.

This work was supported by the Canada Research Chairs programme and the
NSERC Discovery grant.

\appendix
\section{Single particle description \label{app:SP}}

We take the basis $\{ \psi_{A_l}, \psi_{B_u}, \psi_{A_u}, \psi_{B_l}\}$
with $\xi=+1$ in the $K$ valley and the basis $\{\psi_{B_u}, \psi_{A_l},
\psi_{B_l}, \psi_{A_u}\}$ with $\xi=-1$ in the $K'$ valley so that the
single particle nearest neighbour tight binding Hamiltonian for unbiased
bilayer graphene can be written as \cite{mccann-prl96}
\begin{equation}
	H_0 = \left( \begin{array}{cccc}
		0 & 0 & 0 & \xi v_F \pi^\dagger \\
		0 & 0 & \xi v_F \pi & 0 \\
		0 & \xi v_F \pi^\dagger & 0 & \gamma_1 \\
		\xi v_F \pi & 0 & \gamma_1 & 0 
	\end{array} \right), \label{eq:H0def}
\end{equation}
where $\pi = p_x + ip_y$ is the linear expansion of the transfer
integral in the tight binding formalism, $\gamma_1$ is the energy of the
inter-layer dimer bond, and $v_F$ is the Fermi velocity.  
Similarly, the inter-layer bias potential which sets the energy of
lattice sites in the upper layer at $U/2$ and sites in the lower layer
at $-U/2$ causes a gap of magnitude $U$ to open at the $K$ points, and
is described in \cite{mccann-prb74}
\begin{equation}
	H_U = \left( \begin{array}{cccc} \xi U/2 & 0 & 0 & 0 \\
		0 & -\xi U/2 & 0 & 0 \\
		0 & 0 & -\xi U/2 & 0 \\
		0 & 0 & 0 & \xi U/2 
	\end{array} \right).
	\label{eq:HUdef} 
\end{equation}

\subsection{Zero magnetic field \label{app:zerofield}}
In the zero field case, the $\pi$ operators are constructed from the usual
single-particle momentum operators $\pi=-i\hbar \partial_x -\hbar
\partial_y$. The energy spectrum associated with the Hamiltonian $H_0 +
H_U$ is
\begin{equation*}
	E_{\vk,\eta,b} = \nu \Bigg( \frac{\gamma_1^2}{2} +
		\frac{U^2}{4} + \hbar^2 v_F^2 k^2 
	+ b \sqrt{ \frac{\gamma_1^4}{4}
	+ \hbar^2 v_F^2 k^2(U^2 + \gamma_1^2) } \Bigg)^{\frac{1}{2}}
\end{equation*}
The quantum numbers $\nu=\pm1$ and $b=\pm1$ label the band and
branch respectively. Henceforth, we denote the band and valley indices
by the label $\alpha = \{\xi_\alpha, \nu_\alpha, b_\alpha\}$ to
shorten the notation.
The wave functions are given by
\begin{eqnarray*}
	\varphi_{\vk,\alpha} = C_{\vk,\alpha} 
		\frac{e^{i\vk\cdot\vr}}{\sqrt{\mathcal{A}}} 
		\left( \begin{array}{c}
			\frac{\xi_\alpha \hbar v_F k e^{-i\theta_{\vk}}}{\gamma_1}
				\frac{\xi_\alpha \Xi+E_{\vk,\alpha}}{\xi_\alpha \Xi-E_{\vk,\alpha}} f_{\vk,\alpha} \\
			\frac{\xi_\alpha \hbar v_F k e^{i\theta_{\vk}}}{\xi_\alpha \Xi + E_{\vk,\alpha}} \\
			1 \\
			-\frac{\xi_\alpha \Xi+E_{\vk,\alpha}}{\gamma_1} 
				f_{\vk,\alpha} \end{array} \right) \\
	f_{\vk,\alpha} = 
		\frac{\hbar^2 v_F^2 k^2}{(\xi_\alpha \Xi+E_{\vk,\alpha})^2} -1 \\
	\fl
	C_{\vk,\alpha} = \Bigg( 
		\frac{\hbar^2 v_F^2 k^2}{(\xi_\alpha \Xi+E_{\vk,\alpha})^2} +
		1 + 
	\left[\frac{\hbar^2 v_F^2 k^2}{(\xi_\alpha \Xi-E_{\vk,\alpha})^2} + 1
		\right] \frac{(\xi_\alpha \Xi+E_{\vk,\alpha})^2}{\gamma_1^2} 
		f_{\vk,\alpha}^2 \Bigg)^{-\frac{1}{2}}
\end{eqnarray*}
where $\Xi = U/2$ and $\theta_{\vk}$ is the angle of the wave vector
$\vk$ in the graphene plane.

\subsection{Quantizing magnetic field \label{app:finitefield}}

In this case, the $\pi$ operators are constructed from the
guage-invariant momentum in a magnetic field found by making the minimal
coupling substitution in the momentum operator so that $\vp\to \vp + e\vA$.
We label the Landau levels with a set of quantum numbers $\fa =
\{n_\fa, \nu_\fa, \xi_\fa, q_\fa \}$ where $n>0$ is the Landau level
index, $\nu=+1(-1)$ in the conduction (valence) band, $\xi$ is the
valley, and $q$ is the $x$ component of the wave vector which
defines the guiding centre coordinate in the Landau gauge and we assume
that all Landau levels are in the low-energy bands.
Then, the Landau level energies $E_\fa$ are found by
solving a polynomial equation derived using the Landau level operators
$\pi \chi_n =
-i\frac{\hbar\sqrt{2n}}{\lambda_B} \chi_{n-1}$ and $\pi^\dag \chi_n =
i\frac{\hbar\sqrt{2(n+1)}}{\lambda_B} \chi_{n+1}$:
\begin{equation}
	\fl \qquad
	\left[ 2(n_\fa+1) - (\Xi_{\xi_\fa}-E_\fa)^2 \right] 
	\left[ 2n_\fa - (\Xi_{\xi_\fa}+E_\fa)^2 \right] \\
	- \gamma_1^2 (\Xi_{\xi_\fa}^2-E_\fa^2) = 0
	\label{eq:BfieldE}
\end{equation}
where the energies are measured in units of $\hbar v_F/\lambda_B$
and $\Xi_\xi = \xi U/2$. 
The wave function associated with each Landau level is defined by the
index of the Landau function in the third and fourth components. The
$\chi$ functions are only defined for $n_\fa\geq 0$ so components of the
$n_\fa=0$ wave functions which contain indices outside of this range
have zero weight on those components.
The wave functions are:
\begin{eqnarray*}
	\fl
	\varphi_\fa = \frac{d_\fa e^{ik_\fa y}}
		{\sqrt{L\lambda_B\sqrt{\pi}}} 
		\left( \begin{array}{c}
			\frac{-i\xi_\fa \sqrt{2(n_\fa+1)}}
				{\Xi_{\xi_\fa}-E_\fa}
				g_\fa \chi_{n_\fa+1}(r_\fa) \\
			\frac{-i\xi_\fa \sqrt{2n_\fa}}{\Xi_{\xi_\fa} +
			E_\fa} \chi_{n_\fa-1}(r_\fa) \\
			\chi_{n_\fa}(r_\fa) \\
			g_\fa \chi_{n_\fa}(r_\fa)
		\end{array} \right)
		\equiv \frac{d_\fa e^{ik_\fa y}}{\sqrt{L\lambda_B\sqrt{\pi}}} 
		\left( \begin{array}{c} \mathfrak{w}_\fa \chi_{n_\fa+1}(r_\fa) \\
			\mathfrak{x}_\fa \chi_{n_\fa-1}(r_\fa) \\
			\mathfrak{y}_\fa \chi_{n_\fa}(r_\fa) \\
			\mathfrak{z}_\fa \chi_{n_\fa}(r_\fa)
		\end{array} \right) \\
	d_\fa = \left[ \left(
	\frac{2(n_\fa+1)}{(\xi_\fa \Xi - E_\fa)^2} + 1 \right)
	g_\fa^2 + \frac{2n_\fa}{(\xi_\fa \Xi + E_\fa)^2} + 1
	\right]^{-\frac{1}{2}}, \\
	g_\fa = \frac{1}{\gamma_1} \frac{1}{\xi_\fa\Xi + E_\fa}
	\left[ (\xi_\fa \Xi + E_\fa)^2 - 2n_\fa \right],
\end{eqnarray*}
with $r_\fa = x/\lambda_B - q_\fa \lambda_B$ and
\begin{equation*}
	\chi_{n_\fa}(r_\fa) = \frac{1}{\sqrt{n_\fa! 2^{n_\fa}}} 
	e^{-r_\fa^2/2} H_{n_\fa}(r_\fa)
\end{equation*}
In addition to these functions, there are also two `$\delta$-states'
with energy $\pm U/2$ which account for the doubled degeneracy of the
zero energy Landau levels in unbiased bilayer graphene. The wave
functions of these states are
\begin{equation*}
	\varphi_{\delta \fa} = \frac{e^{ik_\fa y}}
		{\sqrt{L\lambda_B\sqrt{\pi}}} 
		\left( \begin{array}{c} \chi_0(r_\fa) \\ 0 \\ 0 \\ 0 \end{array}
		\right)
\end{equation*}

\section{Matrix elements of $\mathcal{F}(\tau)$ \label{app:matelems} }

For convenience, we restate the definition of $\mathcal{F}(\tau) =
\mathcal{H}(\tau) - i\hbar\frac{\partial}{\partial \tau} = H_0 + H_U +
h(\tau) - i\hbar\frac{\partial}{\partial \tau}$.
The matrix elements contain several terms. The terms relating to the
static Hamiltonian and the time derivative are
\begin{equation*}
	\langle\langle j'X' | H_0 + H_U - i\hbar\frac{\partial}{\partial \tau}| 
	jX \rangle \rangle = \left( E_X + j\hbar\Omega_A \right)
	\delta_{j,j'} \delta_{X,X'}
\end{equation*}
The matrix elements of the term associated with the irradiating field
$h(\tau)$ depend on the specific nature of the field and the wave
functions. 
For example, linearly polarized light can be described by the vector
potential $\vA_\mathrm{lin}(\tau) = A \cos(\Omega \tau) \left[ \cos\theta,
\sin\theta \right]$ where $\theta$ is the angle of polarization in the
plane of the graphene with respect to the $x$ axis. This yields
\begin{equation*}
	h_\mathrm{lin}(\tau) = \frac{\xi v_F e F}{\Omega_A} \sigma_x \otimes
	\left( \begin{array}{cc} 0 & e^{i\theta} \\ e^{-i\theta} & 0
	\end{array} \right)
	\cos(\Omega_A\tau).
\end{equation*}
Since we discuss radiation incident at the perpendicular to the graphene
plane, this angle is immaterial and we can substitute $\theta=0$ without
loss of generality.  On the other hand, circularly polarized light is
given by $\vA_\pm(\tau) = A \left[ \cos(\Omega_A\tau), \pm\sin(\Omega_A\tau)
\right]$ where the positive or negative sign corresponds to right- or
left-handed orientations of the polarization:
\begin{equation*}
	h_\pm(\tau) = \frac{\xi v_F e F}{\Omega_A} \sigma_x \otimes
	\left( \begin{array}{cc} 0 & e^{\mp i\Omega_A \tau} \\ 
		e^{\pm i\Omega_A\tau} & 0
	\end{array} \right)
\end{equation*}

\subsection{Zero magnetic field}

In this case the matrix elements of linearly polarized light (taking
$\theta=0$) are zero for states with unequal wave vectors, and as follows
for states in the same valley and with the same spin, which also have
identical wave vector $\vk$: 
\begin{eqnarray*}
	\fl
	\langle\langle j'\vk \alpha' | h_\mathrm{lin}(\tau) | 
		j\vk \alpha \rangle\rangle
	= \frac{\hbar v_F^2 k e F}{2\Omega} 
		\left( \delta_{j',j+1} + \delta_{j',j-1} \right) 
		C_\alpha C_{\alpha'} \delta_{\xi,\xi_\alpha} 
			\delta_{\xi,\xi_{\alpha'}} \times \\
	\fl
	\quad
	\times \left[ \frac{e^{-i\theta_{\vk}}}{\xi \Xi + E_{\alpha'}} +
		\frac{e^{i\theta_{\vk}}}{ \xi\Xi + E_\alpha } - 
		\frac{f_\alpha f_{\alpha'}}{\gamma_1^2} 
		\left(\xi\Xi + E_{\alpha'}\right)
		\left(\xi\Xi + E_\alpha\right) 
		\left( \frac{e^{i\theta_{\vk}}}{\xi\Xi - E_{\alpha'}} 
			+ \frac{e^{-i\theta_{\vk}}}{\xi\Xi - E_\alpha} \right) \right].
\end{eqnarray*}
The matrix elements of states with different spin or valley are zero.
Similarly, the matrix elements of the Hamilonian associated with
circularly polarized radiation are diagonal in the wave vector, spin,
and valley and are given by 
\begin{eqnarray*}
	\fl
	\langle\langle j'\vk X' | h_\pm(\tau) | j\vk X \rangle\rangle
		= \frac{\hbar v_F^2 k e F}{\Omega} C_X C_{X'} 
		\delta_{\xi,\xi_X} \delta_{\xi,\xi_{X'}} \times \\
	\fl
	\qquad\qquad 
	\times\bigg\{ \delta_{j',j\mp1} e^{i\theta_{\vk}} 
		\left[ \frac{1}{\xi\Xi + E_\alpha} - 
		\frac{f_\alpha f_{\alpha'}}{\gamma_1^2}
		\frac{(\xi \Xi + E_{\alpha'})(\xi\Xi + E_\alpha)}
		{\xi \Xi - E_{\alpha'}} \right] + \\
	\fl
	\qquad\qquad\qquad\qquad
	+ \delta_{j', j\pm1} e^{-i\theta_{\vk}} 
		\left[ \frac{1}{\xi \Xi + E_{\alpha'}} - 
		\frac{f_\alpha f_{\alpha'}}{\gamma_1^2} 
		\frac{(\xi\Xi + E_\alpha)(\xi\Xi+E_{\alpha'})}{\xi\Xi - E_\alpha}
		\right]
	\bigg\}.
\end{eqnarray*}

\subsection{Quantizing magnetic field}

The matrix elements in this case are
\begin{eqnarray*}
	\fl
	\langle\langle j'\fa' | h_\mathrm{lin}(\tau) | j\fa \rangle\rangle
		= \frac{iv_FeF}{\sqrt{2}\Omega} 
		\left( \delta_{j',j+1} + \delta_{j',j-1} \right) 
		d_\fa d_{\fa'} \delta_{k_\fa, k_{\fa'}}
		\delta_{\xi,\xi_\fa} \delta_{\xi,\xi_{\fa'}}
		\times \\
	\fl
	\qquad \qquad 
	\times \bigg\{ \delta_{n_\fa,n_{\fa'}+1} \sqrt{n_\fa} 
		\left( \frac{g_\fa g_{\fa'}}{\xi\Xi - E_{\fa'}} 
		- \frac{1}{\xi\Xi + E_\fa} \right) \\
	\fl
	\qquad\qquad\qquad\qquad
		- \delta_{n_\fa,n_{\fa'}-1} \sqrt{n_\fa+1}
		\left( \frac{g_\fa g_{\fa'}}{\xi\Xi - E_\fa} 
		- \frac{1}{\xi\Xi + E_{\fa'}} \right) \bigg\},
\end{eqnarray*}
and
\begin{eqnarray*}
	\fl
	\langle\langle j'\fa' | h_\pm(\tau) | j\fa \rangle\rangle
	= \frac{\sqrt{2}iveF}{\Omega} d_\fa d_{\fa'} 
		\delta_{k_\fa,k_{\fa'}}
		\delta_{\xi,\xi_\fa} \delta_{\xi,\xi_{\fa'}} \times \\
	\times \Big\{ 
		\delta_{j',j\mp1} \delta_{n_\fa',n_{\fa-1}} \sqrt{n_\fa}
		\left( \frac{g_\fa g_{\fa'}}{\xi\Xi - E_{\fa'}} 
		- \frac{1}{\xi\Xi + E_\fa} \right) \\
	\qquad\qquad
		- \delta_{j',j\pm1} \delta_{n_{\fa'}, n_{\fa+1}} \sqrt{n_\fa+1}
		\left( \frac{g_\fa g_{\fa'}}{\xi\Xi - E_\fa} - 
		\frac{1}{\xi\Xi + E_{\fa'}} \right) \Big\}.
\end{eqnarray*}


\end{document}